# Breakdown of self-cleaning mechanism for nanoscale interfacial substances in tiny-angle twisted bilayer graphene

Chao Yan, Ya-Xin Zhao, Yi-Wen Liu, Lin He[†]
Center for Advanced Quantum Studies, Department of Physics, Beijing Normal University, Beijing, 100875, People's Republic of China

[†]Correspondence and requests for materials should be addressed to Lin He (e-mail: helin@bnu.edu.cn).

**Realization of high-quality van der Waals (vdW) heterostructures with tailored properties by stacking two-dimensional (2D) layers requires atomically clean interfaces. Because of strong adhesion between the constituent layers, the vdW forces could drive trapped contaminants together into submicron-size 'bubbles', which leaves large interfacial areas atomically clean. Such a phenomenon is dubbed "self-cleaning" mechanism in 2D systems. Here, we demonstrate the breakdown of self-cleaning mechanism for nanoscale interfacial bubbles in tiny-angle twisted bilayer graphene (TBG). In the tiny-angle TBG, there is a triangular network of domain boundaries due to structural reconstruction. Our experiments indicate that the bubbles will mainly move along the triangular network of domain boundaries when the sizes of the bubbles are comparable to that of an AA-stacking region in the TBG. When the size of the bubble is smaller than that of an AA-stacking region, the bubble becomes motionless and is fixed in the AA-stacking region because of its large out-of-plane corrugation. Our results reveal a substantial influence of the moiré superlattice on the motion of nanoscale interfacial substances.**

Two-dimensional (2D) van der Waals (vdWs) materials provide a platform that allows creation of heterostructures and homostructures with tailored electronic, mechanical, thermal, optical and optoelectronic properties[1–6]. The realization of these attractive properties requires high quality devices with interfacial areas atomically clean and free of contamination. However, it is almost inevitable to leave contaminants at the interface in the process of fabricating vdWs devices[7–15]. Therefore, researchers have made great efforts to optimize sample-fabricated methods to continuously improve the quality of 2D devices[1–6,16–19]. Thanks to the vdWs force between the constituent layers, the interfacial substances are very mobile and they can be merging into submicron-size 'bubbles'. Such a phenomenon is dubbed "self-cleaning" mechanism in 2D systems because that it leaves large interfacial areas atomically sharp and free of contaminations[7–15]. In 2D vdW heterojunctions and homojunctions, the interlayer interaction introduces a moiré-periodic network of out-of-plane corrugations[20–36]. However, effects of such corrugations on the motion of the interfacial substances are completely ignored and have not been addressed up to now. In this work, motion of nanoscale interfacial bubbles in tiny-angle twisted bilayer graphene (TBG) has been investigated using scanning tunneling microscopy (STM). Our experiment demonstrates that the moiré-periodic network of corrugations in the TBG has a substantial influence on the motion of the nanoscale interfacial bubbles. When the size of the bubble is comparable to that of an AA-stacking region in the TBG, it will mainly move along the moiré-periodic network of domain boundaries. When the size of the bubble is smaller than that of an AA-stacking region, the bubble becomes motionless and is fixed in the AA-stacking region, indicating breakdown of the self-cleaning mechanism.

Figure 1(a) shows schematic of our experimental device set-up and the TBG samples are obtained by wet transfer technology of graphene layer by layer on mechanical-exfoliated $WSe_2$ sheets (see Supplemental Figs. S1 and S2 for details)[24,37-39]. Figure 1(b) shows a representative STM image of the obtained TBG. We can clearly identify the moiré superlattice from the periodic corrugations in the image and the twist angles can be obtained as about 0.5° based on the measured periods $D$ according to

$D = a/[2\sin(\theta/2)]$, where $a \approx 0.246$ nm. For such a tiny-angle TBG, there is a strain-accompanied lattice reconstruction that results in large triangular stacking domains (the *AB* and *BA* regions) with a triangular network of domain walls (DWs)[21,23–25,29–36]. Simultaneously, the *AA* stacking regions are reduced to minimize the total energy of the system (here the radius of the *AA* stacking region $R_{AA}$ is about 3 nm). The representative scanning tunneling spectroscopy (STS) spectra recorded in the *AA*, *AB*(*BA*), and *DW* regions are shown in Fig. 1(c). In both the *AA* and *DW* regions, the nearly flat bands arising from the *AA* modes and the *DW* modes, as reported previously[24,31], are observed as pronounced peaks in the STS spectra (see Supplemental Fig. S3 for more STM characterizations of the TBG). These measurements indicate that the interface of the studied region is atomically sharp and free of contaminations. However, the interfacial contamination (such as adsorbed water and hydrocarbons used in the transfer process) is inevitable present in the TBG and Fig. 1(d) shows a representative STM image of the TBG with interfacial contaminations. Besides the periodic moiré superlattice, two additional nanoscale bubbles can be clearly identified (see Supplemental Figs. S4, S5 and S6 for more STM characterizations of the interfacial contamination in the TBG). Previously, a universal scaling law of $h/R \approx 0.12$ is found experimentally for large bubbles ($R > 50$ nm) confined by graphene, where $h$ is the maximum height and $R$ is the base radius of the graphene bubble[9]. Such a behavior can be well described based on the theory of elasticity of membranes. In our experiment, the size of the bubbles is much smaller than 50 nm and the shape of the bubbles, obviously, deviates from the universal scaling law, as shown in Figs. 1(e) and 1(f). This behavior is reasonable because that the shape of nanoscale bubbles with $R < 50$ nm is beyond the description of the simple membrane theory[9,10]. Moreover, the trapped substance of the nanobubble[10] and the structure reconstruction of the TBG[20–36] can affect the shape of the interfacial nanobubble.

An interesting observation is that the movement of the nanobubble can be actuated by STM tip scanning and the nanobubble's trajectory can be traced in the STM measurements, as summarized in Figs. 2(a)-2(d) (the radius of the nanobubble is about 11.8 nm). The white arrow (Figure 2(b)) points to the direction of movement of the bubble, which is consistent with the direction of movement of the STM tip. Although the motion of the nanobubble is too fast to be caught in the STM measurement, it leaves traceable trajectory in the TBG, as shown in Figs. 2(b) and 2(d). According to previous studies[9,10,40], the pressure inside the nanoscale bubble, which is determined by the adhesion forces between the layers, can reach about 1 GPa. Such a large variation of the pressure in the moving process of the bubble will definitely affect the local stacking configuration of the TBG. Therefore, the networks of moiré pattern in the region around the trajectory become distorted and even undetectable, as shown in Fig. 2(b). It is interesting to find out that the networks of moiré pattern recover themselves to a contamination-free structure after a few hundred seconds, as shown in Fig. 2(c). Such a long-time recovery of the interface, which is important for the tracing of the motion of the nanobubble, is consistent with the extremely long-period oscillations of the interlayer separation observed recently in the TBG[41]. The most striking observation in our experiment is the size dependent motion of the nanobubble in response to the mechanical stimuli of the STM tip, as shown in Figs. 2(e)-2(h) as an example. The larger nanobubble with a radius of 11.2 nm is very mobile and can be easily actuated by the STM tip scanning, leaving a traceable trajectory in the TBG, as shown in Fig. 2(f). Similarly, the interface of the TBG around the trajectory recovers after a few hundred seconds of the movement (Fig. 2(g)). However, the two smaller bubbles with radii of 4.2 nm and 4.4 nm respectively, as circled by red circles in Figs. 2(e)-2(g), are motionless and seem to be pinned in the AA-stacking regions, no matter how we change the scanning conditions. Here we should point out that the size of the bubbles is difficult to be measured exactly because of the contribution of corrugation of the TBG. This is especially the case for the two smaller bubbles. The radius of the AA-stacking regions with trapped nanobubble is only slightly larger than that without trapped nanobubble, indicating that the real size of the small nanobubbles is much smaller than the size of the AA-stacking region. For simplicity, here we use the measured radius, including the contribution of corrugation of the TBG, as the nominal radius of the nanobubbles. The above result reveals that the nanobubbles cannot coalesce into a large bubble in the TBG when the radius is smaller than the size of the AA-stacking region, indicating breakdown of the self-cleaning mechanism in this case.

To further explore effects of the size on the motion of the nanobubble in the TBG, we use the STM tip to *in situ* tune the size of the nanobubble. By reducing the distance between the STM tip and the graphene, a large nanobubble with a radius of 10.8 nm can be divided into two intermediate-size nanobubbles with radii of 7.8 nm and 9.3 nm respectively, as shown in Fig. 3(a) (Here we also use the measured radius, including the contribution of corrugations of TBG, as the nominal radius of the nanobubbles). They move in opposite directions due to the conservation of momentum. It is interesting to find out that the motion of the two intermediate-size nanobubbles in response to the tip scanning is distinct from that of the large nanobubbles. For the large nanobubble, each mechanical stimulus of the STM tip can cause a movement of the nanobubble over several hundreds of nanometers and the moving can be in any selected direction. However, for the intermediate-size nanobubble, each movement induced by the tip scanning usually ranges from several nanometers to several tens of nanometers and, more strikingly, the motion can only be along the network of the DWs, as shown in Figs. 3(b)-3(d) (see Supplemental Fig. S8 for more experimental data). The size dependent motion of the nanobubble is further confirmed by continuously increasing the size of the nanobubble. As summarized in Figs. 3(e)-3(h), the nanobubble increases from an intermediate one to a large one through coalescence of interfacial substances in the moving process. After reaching a "critical" size, the motion of the bubble become the same as the large nanobubble, as shown in Fig. 3(g).

Obviously, the interfacial nanobubbles in the TBG, depending on their sizes, exhibit three distinct behaviors in response to the mechanical stimuli of the STM tip. The large nanobubbles are very mobile and can move very fast in any selected direction, whereas, the small nanobubbles are motionless and are pinned in the AA-stacking regions. The intermediate-size nanobubble only moves along the network of the DWs in the TBG. According to our experimental results, the "critical" radii to divide the three regimes of the nanobubbles are determined as $1.5R_{AA}$ and $3.4R_{AA}$ (here we use the nominal radii measured by STM), as shown in Fig. 4(a). Such a result demonstrates explicitly that the moiré-periodic network of corrugations in the TBG has a substantial influence on the motion of the nanobubbles. In the structure-reconstructed TBG, the interlayer distances in the different stacking regions are quite different and $D_{AA} > D_{DW} > D_{AB}$ ($\approx D_{BA}$)[20-36], as schematically shown in Figs. 4(b) and 4(c). This provides the moiré-periodic network of local minima and paths to affect the motion of the interfacial nanobubbles. In the scanning process of the STM, the tip can locally lift the topmost graphene beneath the tip to introduce mechanical stimuli to the nanobubbles. For the large nanobubbles with radii much larger than the $R_{AA}$, the effect of out-of-plane corrugations in the TBG on their motion is negligible. However, when the size of the nanobubble is comparable to the $R_{AA}$ (i.e., the intermediate-size nanobubble), its motion should be strongly affected by the moiré-periodic out-of-plane corrugations and the trajectory of the nanobubbles is expected to be limited to along the triangular networks of the DWs. For the case that the size of the nanobubble is much smaller than the $R_{AA}$, the nanobubble is strongly trapped in the AA-stacking region due to the large out-of-plane corrugations and the topmost graphene fails to stimulate the motion of the nanobubble from the local minimum. Then the nanobubble becomes motionless and is pinned in the AA-stacking region.

In summary, we study the motion of nanoscale interfacial bubbles in tiny-angle TBG in response to the STM tip. Depending on the sizes of the nanobubbles, the interfacial nanobubbles in the TBG exhibit three distinct behaviors in response to the mechanical stimuli of the STM tip. Our experiment demonstrates that the motion of the nanoscale interfacial bubbles is strongly affected by the moiré-periodic network of corrugations in the TBG and reveals the breakdown of the self-cleaning mechanism when the size of the nanobubble is smaller than that of an AA-stacking region.

**Acknowledgments**

This work was supported by the National Key R and D Program of China (Grant Nos. 2021YFA1401900, 2021YFA1400100) and National Natural Science Foundation of China (Grant Nos. 12141401,11974050).

**Author contributions**

C.Y. fabricated the samples and performed STM experiments with the contributions of Y.W.L.. C.Y., Y.W.L., Y.X.Z. and L.H. analyzed the data. L.H. conceived and provided advice on the experiment, and analysis. L.H. and C.Y. wrote the paper. All authors participated in the data discussion.

**Notes**

The authors declare no competing financial interests.

**Methods**

**CVD Growth of aligned Graphene.** We use low-pressure chemical vapor deposition (CVD) method to grow aligned graphene on copper foils (0.025 mm, 99.8% purity, purchased from Alfa Aesar). The growth process is divided into four main stages, as illustrated in Supplementary Figure S1. In the first stage, the copper foil was suspended on a quartz support and heated to 1035°C with $H_2$ (50 sccm) and Ar (100 sccm). In the second stage, the copper foil was heated for 10 hours to turn into a single crystal with $H_2$ (50 sccm) and Ar (100 sccm). In the third stage, keeping the flow rates of Ar and $H_2$ constant, we introduced 5 sccm methane to grow graphene. Finally, the copper foil was slowly cooled down to room temperature in 120 min.

**Sample preparation.** The copper coated with aligned graphene was divided into two pieces. One of the copper foils was spun coated with polymethyl methacrylate (PMMA). Ammonium persulfate solution was used to etch the copper foil to release the graphene on PMMA film. The other copper foil was used to scoop up the PMMA film. Meanwhile, the twisted angle is strictly controlled by controlling the relative twisted angle of PMMA and copper foil. Then, the sample was placed in a nitrogen environment for 12 h to couple the two layers of graphene together. Subsequently, the copper foil was etched by ammonium persulfate solution, by this time the twisted bilayer graphene has been attached to the PMMA. After thorough cleaning with deionized water, the PMMA film was transfer to Si/SiO$_2$ covered with a thin layer of WSe$_2$ (purchased from Shanghai Onway Technology Co., Ltd) which was exfoliated from WSe$_2$ bulk. In order to better carry out the experiment, we do not do any heating treatment on the sample. Finally, we used acetone to remove the PMMA covered on the sample surface. So far, we have successfully fabricated our sample.

**STM and STS Measurements.** The STM system was an ultrahigh vacuum (~10$^{-11}$ Torr) single probe scanning probe microscope USM-1500 from UNISOKU. All the STM experiments performed at liquid-nitrogen temperature ~77 K with constant-current scanning mode. The STS spectrum, i.e. the *dI / dV–V* curve, was carried out with a standard lock-in technique by applying alternating current modulation of the bias voltage of 5 mV (871 Hz) to the tunneling bias. The scanning tunneling microscope tips were obtained from electrochemical corrosion of tungsten wire.

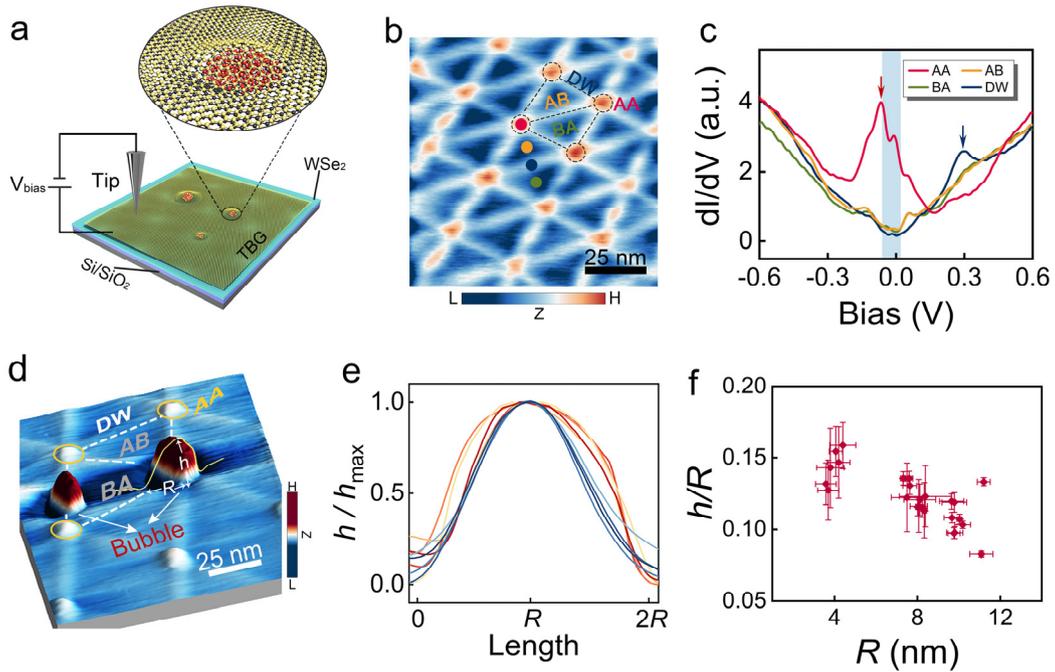

**Figure 1. Interfacial nanobubbles in tiny-angle TBG.** (a) Schematic of the experimental device set-up. (b) A representative STM image of the TBG ($V_b$ = 0.6 V, $I$ = 0.2 nA). The *AA*, *AB*, and *DW* stacking orders in the TBG are marked on the corresponding regions. (c) Representative STS spectra measured at *AA*, *AB*, *BA* and *DW* regions, as labelled in panel (b). The flat bands in the *AA* region and *DW* modes are marked by red arrow and blue arrow respectively. The helical network of the TBG can be observed within the energy window of the light blue rectangle. (d) A representative STM topography image of TBG with two nanobubbles trapped between the two adjacent layers ($V_b$ = 1 V, $I$ = 0.2 nA). The orange circles circled are *AA* stacking regions of the TBG. The white dotted lines connecting *AA* stacking are DWs. The surrounding blue regions separated by DWs are *AB* and *BA* stacking regions. The yellow line displays the contour of a nanobubble with height $h$ and radius $R$. The scale bar is 25 nm. (e) Scaled and normalized cross-sectional profiles of several interfacial nanobubbles. The profiles of different bubbles marked by different colors. (f) Measured aspect ratios of different nanobubbles trapped in the TBG.

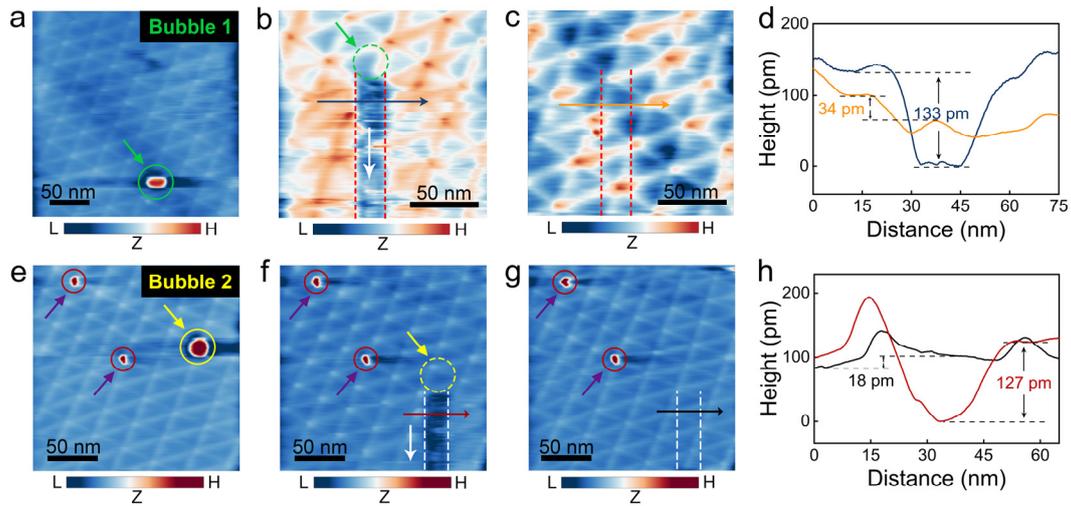

**Figure 2. Motion of nanobubbles in response to tip scanning.** (a) STM topography image of Bubble 1, marked by green circle, confined in the TBG. (b) Trajectory of the Bubble 1 in response to STM tip scanning from the top to the bottom. The trajectory, as marked by dotted red lines, can be observed due to the movement of the Bubble 1. The dotted green circle is the original location of the nanobubble shown in panel (a) and the white arrow points to the direction of its movement. (c) After about 700 seconds, the trajectory disappears and the moiré superlattice recovers in the STM image. (d) Height profiles along the brown arrow and the blue arrow marked in (b) and (c), respectively. (e) STM topography image of three nanobubbles confined in the TBG. Nanobubbles with smaller radii, as circled by red circles, are fixed in the AA stacking regions. (f) Movement of Bubble 2, marked by yellow circle, in response to STM tip scanning from the top to the bottom. The yellow dotted circle indicates the initial location of Bubble 2. The white dashed lines are guides for the eye to show the trajectory of Bubble 1. The white arrow indicates the moving direction of Bubble 2. (g) After about 900 seconds, the trajectory disappears and the moiré superlattice recovers in the STM image. (h) Height profiles along the red arrow and the black arrow marked in (f) and (g), respectively. (a)-(c) $V_b$ = 1 V, $I$ = 0.2 nA. (e)-(g) $V_b$ = 0.8 V, $I$ = 0.2 nA.

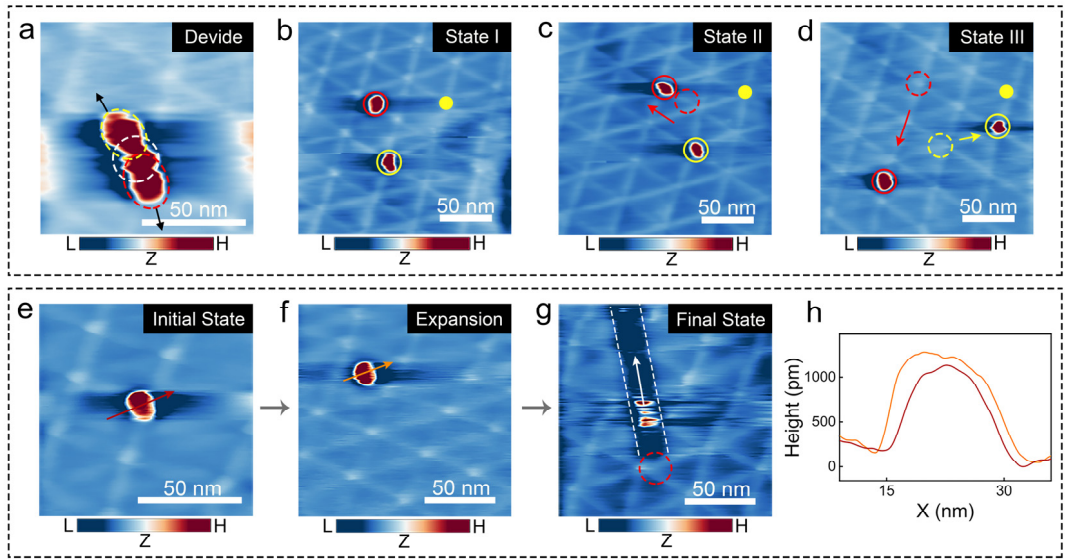

**Figure 3. Manipulating nanobubbles with the STM tip.** (a) A nanobubble with a larger radius is divided into two smaller nanobubbles by using the STM tip. The white dotted circle indicates the nanobubble's original location. The yellow dotted ellipse and red dotted ellipse represent the two new nanobubbles, respectively. The black arrows are guides to the eye, indicating the moving directions of the two nanobubbles ($V_b$ = 1.7 V, $I$ = 0.2 nA). (b)-(d) STM topography images recording the motion of the two nanobubbles. The yellow dot marks the same position in the TBG to clearly show the relative positions of the nanobubbles in the three STM images. The dotted circles indicate the initial locations of the nanobubbles. And the arrows indicate the directions of motion of the nanobubbles ($V_b$ = 0.8 V, $I$ = 0.2 nA). (e)-(g) STM topography images showing that the size of a nanobubble increases gradually by merging interfacial substance. The motion of the nanobubble depends on its size. (e) $V_b$ = 0.8 V, $I$ = 0.05 nA; (f) $V_b$ = 1.3 V, $I$ = 0.1 nA; (g) $V_b$ = 1 V, $I$ = 0.1 nA. (h) Cross-sectional profiles of nanobubbles as arrows marked in (e) and (f).

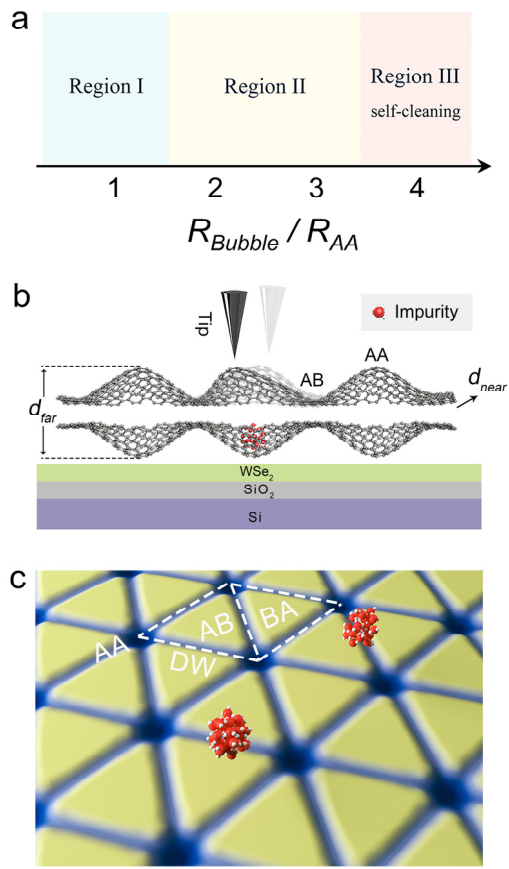

**Figure 4. Size-dependent motion of the nanobubbles in the TBG.** (a) Three regimes are divided due to the size-dependent motion of the nanobubbles in the TBG. Region I: bubbles are pinned in the *AA*-stacking regions. Region II: bubbles move along the network of the DWs. Region III: bubbles can move along any direction. (b) Schematic of the corrugations of the TBG. The STM tip can locally introduce mechanical stimuli to the interfacial contaminant through the topmost graphene layer. (c) Schematic of out-of-plane corrugations in the bottom graphene layer of the TBG. The networks are the DWs, which provide the moving paths for the bubbles with intermediate size.